\documentclass{camera}
\usepackage{graphicx}  
\usepackage{color}
\usepackage[colorlinks=true]{hyperref}

\definecolor{dgreen}{cmyk}{1.,0.,1.,0.4}        
\definecolor{orange}{cmyk}{0.,0.353,1.,0.}    

\usepackage{lineno}

\begin{document}


%
\title{Is hadronic flow produced in p--Pb collisions at the Large Hadron Collider?}

%
\author{You Zhou$^{~a,}\footnote{~you.zhou@cern.ch}$, Xiangrong Zhu$^{~b}$, Pengfei Li$^{~b}$, Huichao Song$^{~b}$}
\organization{$^{a~}$ Niels Bohr Institute, University of Copenhagen, Denmark\\
$^{b}$ Department of Physics and State Key Laboratory of Nuclear Physics and Technology, Peking University, China
}

\maketitle

\begin{abstract}

Using the Ultra-relativistic Quantum Molecular Dynamics ({\tt UrQMD}) model, we investigate the azimuthal correlations in p--Pb collisions at $\sqrt{s_{_{\rm NN}}}=5.02$ TeV. It is shown that the simulated hadronic p--Pb system can not generate the collective flow signatures, but mainly behaves as a non-flow dominant system. However, the characteristic $v_{2}(p_{\rm T})$ mass-ordering of pions, kaons and protons is observed in {\tt UrQMD} simulations, which is the consequence of hadronic interactions and not necessarily associated with strong fluid-like expansions.

\end{abstract}


The relativistic heavy ion collisions at the Large Hadron Collider (LHC) have provided strong evidences for the creation of the Quark--Gluon Plasma (QGP). One of the crucial observables is the azimuthal anisotropy of the transverse momentum distribution for produced hadrons~\cite{Ollitrault:1992bk}, usually characterized by the Fourier flow-coefficients~\cite{Voloshin:1994mz}. The second Fourier flow-coefficient $v_{2}$ is called elliptic flow, which has been systematically measured and studied at the LHC~\cite{Aamodt:2010pa}. These results, together with the comparisons to theoretical model calculations, provides important information on the Equation of State (EoS) and the transport properties of the QGP.
The azimuthal correlations in $\sqrt{s_{_{\rm NN}}} =$ 5.02 TeV p--Pb collisions were also measured at the LHC with the original purposes of providing reference data for the high energy Pb--Pb collisions. However, a large amount of unexpected collective behaviors have been discovered~\cite{ABELEV:2013wsa} and the results can be semi-quantitatively described by (3+1)-d hydrodynamic simulations~\cite{Bozek:2013ska, Werner:2013ipa}. It might suggest that the large collective flow has been developed in the small p--Pb systems.
In this study~\cite{Zhou:2015iba}, we utilize a hadron cascade model Ultra-relativistic Quantum Molecular Dynamics ({\tt UrQMD} version 3.4)~\cite{Bass:1998ca} to simulate the evolution of the hadronic matter and then study the azimuthal correlations of final produced hadrons.

\begin{figure}
\centering
\includegraphics[width=0.45\textwidth]{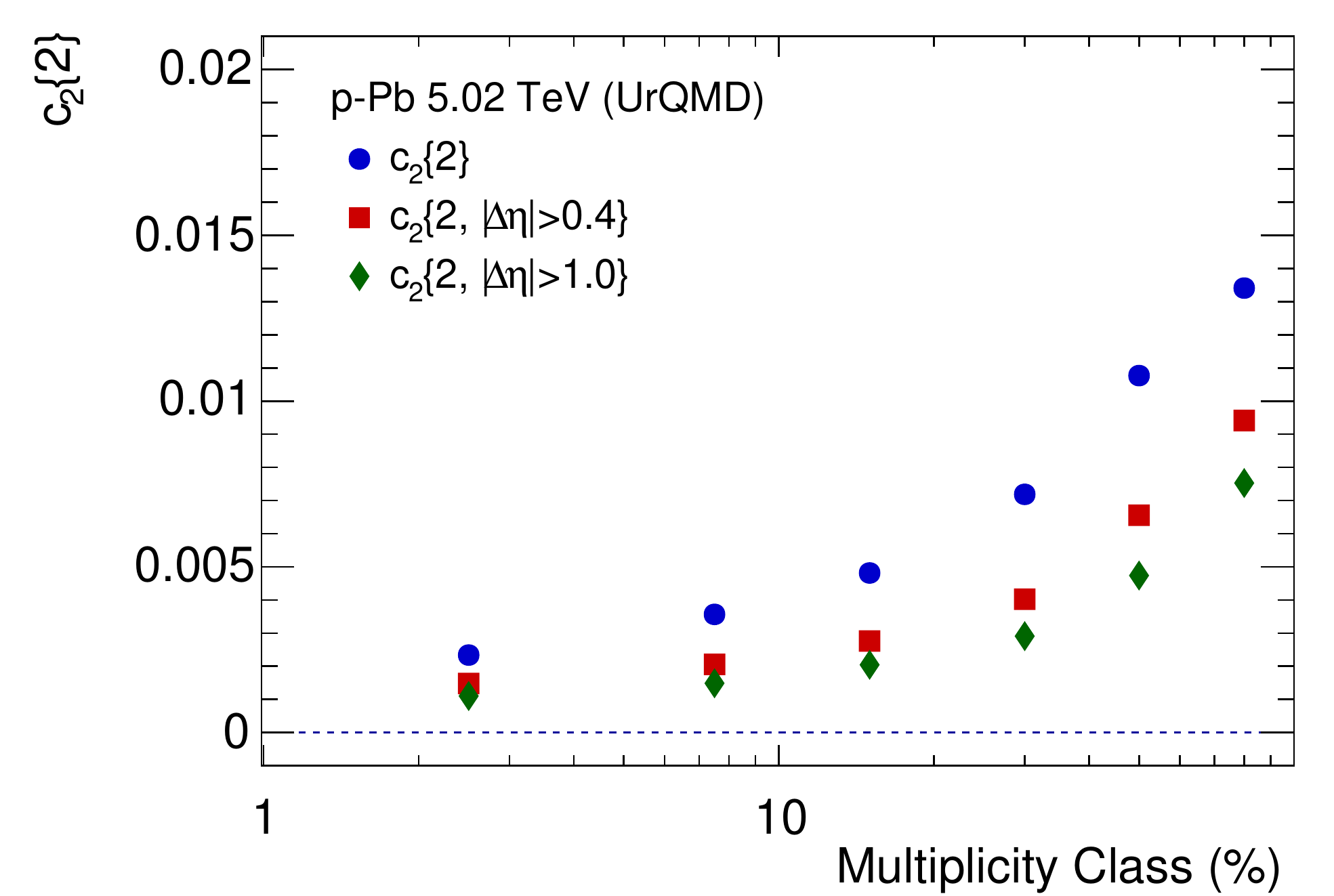}
\includegraphics[width=0.48\textwidth]{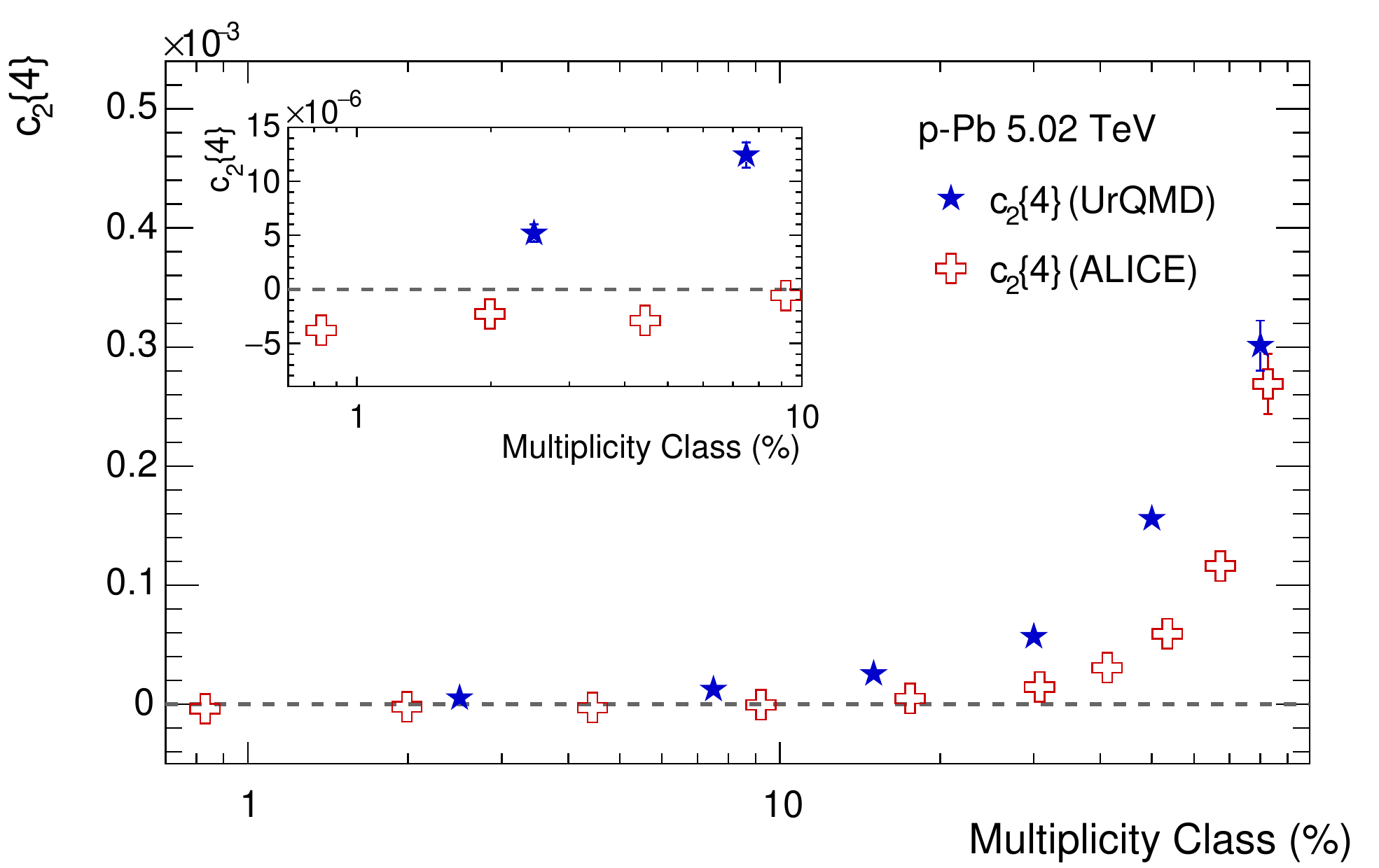}
\caption{Multiplicity class dependence of $c_{2}\{2\}$ (left) and $c_{2}\{4\}$ (right).}
\label{figure:c224} 
\end{figure}

Figure~\ref{figure:c224} (left) presents the centrality dependence of the 2-particle cumulant $c_{2}\{2\}$ from {\tt UrQMD} model. With various pseudorapidity gaps, $c_{2}\{2\}$ exhibits decreasing trends from low multiplicity events to high multiplicity events. As the pseudorapidity gap increases, the magnitudes of $c_{2}\{2\}$ become weaker for {\tt UrQMD}. These results agree with the suppression of the so-called non-flow effects, which is the azimuthal correlations not associated with the symmetry plane. When the pseudorapidity gap $|\Delta \eta|$ is larger than 1.0, $c_{2}\{2\}$ from {\tt UrQMD} still presents strong centrality dependence, showing a typical non-flow behavior. It indicates that {\tt UrQMD} hadronic expansion could not generate enough flow in a small p--Pb system, and non-flow effects still dominate the 2-particle correlations even with $|\Delta \eta| >$ 1.0.
To better understand the hadronic systems simulated by {\tt UrQMD}, we investigate the 4-particle cumulant of the second Fourier flow-coefficient $c_{2}\{4\}$, which is less sensitive to non-flow effects. Figure~\ref{figure:c224} (right) shows that the $c_{2}\{4\}$ measurement exhibits a transition from positive to negative values for the most central collisions, indicating the creation of flow-dominated systems in the high multiplicity events. However, $c_{2}\{4\}$ calculations from {\tt UrQMD} simulations keep positive for all available multiplicity classes, including central collisions. As a result, real values of $v_{2}\{4\}$ can not be extracted. This comparison further illustrates the difference between the p--Pb systems created in experiment and simulated by {\tt UrQMD}. Without the contributions from the initial stage and/or the QGP phase, the measured flow-like 4-particle correlations in high multiplicity events can not be reproduced by a microscopic transport model with only hadronic scatterings and decays.

\begin{figure}
\centering
\includegraphics[width=0.48\textwidth]{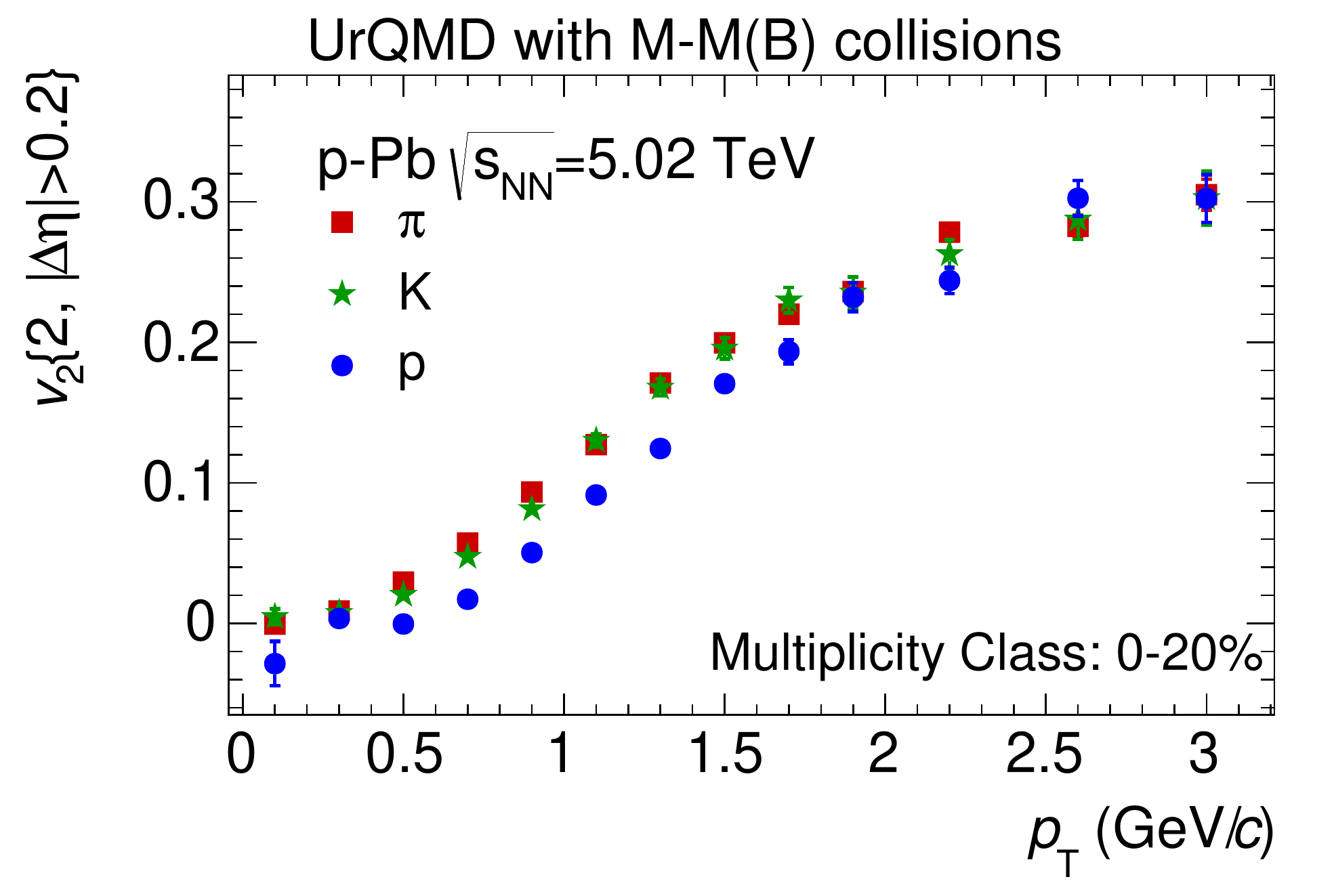}
\includegraphics[width=0.48\textwidth]{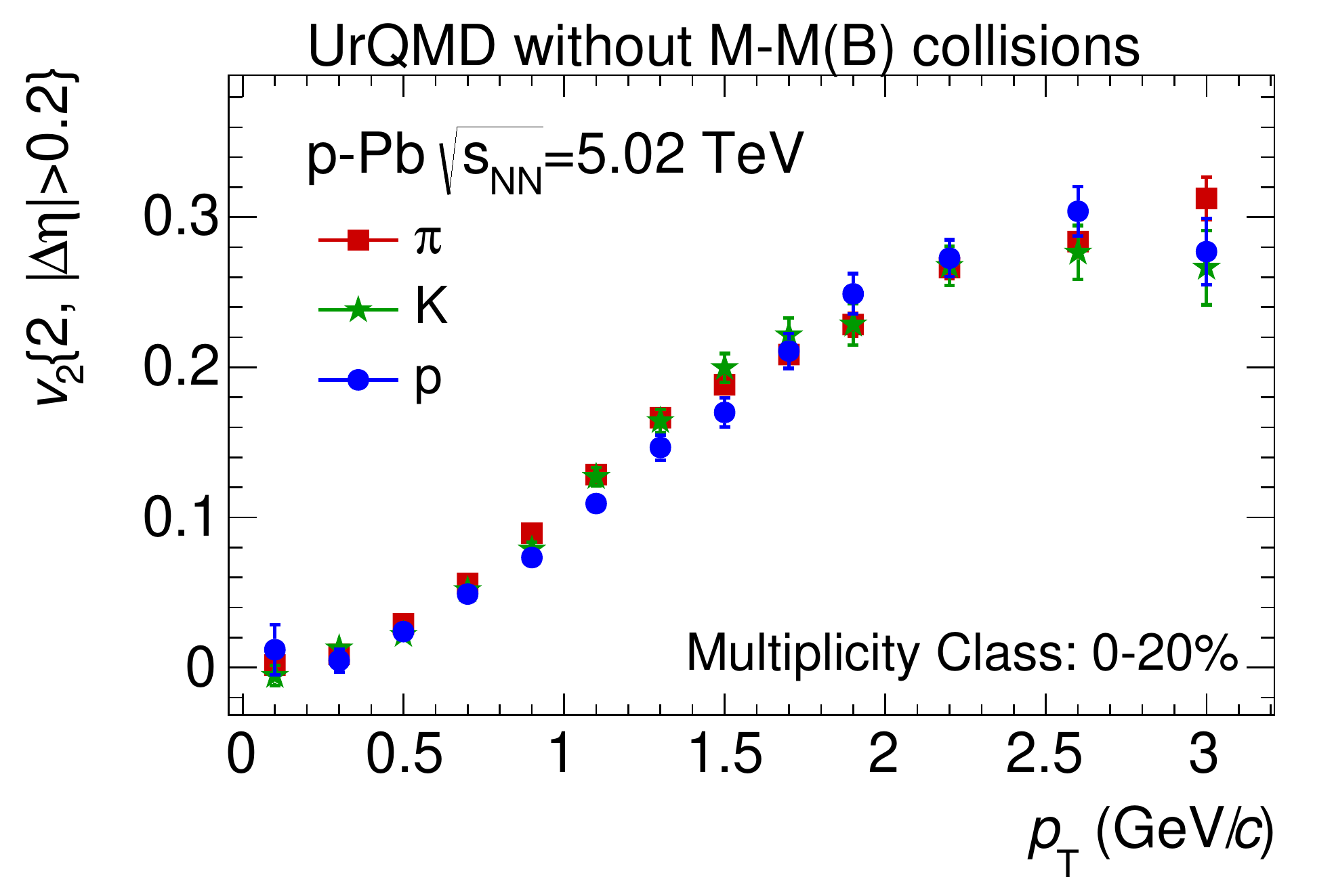}
\caption{$v_{2}(\it{p}_{\rm T})$ of pions, kaons and protons in p--Pb collisions at $\sqrt{s_{_{\rm NN}}} =$ 5.02 TeV, calculated from {\tt UrQMD} with (left) and without (right) M-M and M-B collisions.}
\label{fig:pidv2Gap02} 
\end{figure}

In addition, the characteristic feature of $v_{2}(p_{\rm T})$ mass-ordering among pions, kaons and protons was observed in p--Pb collisions~\cite{ABELEV:2013wsa}. This mass ordering feature is similar with what observed in Pb--Pb collisions~\cite{Abelev:2014pua} and can be roughly reproduced by hydrodynamic calculations~\cite{Bozek:2013ska, Werner:2013ipa}. It was token as a strong evidence for the collective expansion of the p--Pb systems created in $\sqrt{s_{_{\rm NN}}}=$ 5.02 TeV collisions.
Figure~\ref{fig:pidv2Gap02} shows that a clear $v_2$ mass-ordering among pions, kaons and protons below 2 GeV is also observed in {\tt UrQMD}. Such mass-ordering pattern, caused by only hadronic interactions, qualitatively agrees with the ALICE measurement~\cite{ABELEV:2013wsa}.
In {\tt UrQMD}, the unknown hadronic cross sections are calculated by the additive quark model ({\tt AQM}) through counting the number of constituent quarks within two colliding hadrons. As a result, the main meson-baryon (M-B) cross sections from {\tt AQM} are about 50\% larger than the meson-meson (M-M) cross sections, leading to the  $v_{2}$ splitting between mesons and baryons after the evolution of hadronic matter. It is also observed in Fig.~\ref{fig:pidv2Gap02} (right) that when switch off the M-B and M-M interaction channels in {\tt UrQMD}, the $v_{2}$ mass-ordering almost disappears. The comparison of Fig.~\ref{fig:pidv2Gap02} (left) and (right) illustrates that the hadronic interactions could lead to a $v_{2}$ mass-ordering feature, even for small p--Pb systems without enough flow generation.

In summary, using {\tt UrQMD} hadron cascade model, we studied azimuthal correlations in p--Pb collisions at $\sqrt{s_{_{\rm NN}}} =$ 5.02 TeV. It is found that hadronic interactions alone could not generate sufficient collective flow as observed in experiment. In order to fit the azimuthal correlations measurements, the contributions from the initial stage and/or the QGP phase can not be neglected.
In addition, we extended our study of azimuthal correlations to identified hadrons. A $v_2$ mass-ordering was generated by {\tt UrQMD}, which is similar to the ALICE measurements. This characterize feature is mainly caused by hadronic interactions. The experimentally observed $v_{2}$ mass-ordering alone is not necessarily explained as a flow signal associated with the strong fluid-like expansions.


%
\end{document}